\documentclass[longbibliography,physrev,aps,twocolumn]{revtex4-2}
\usepackage{graphicx,epsfig}
\usepackage{times}
\usepackage{amsmath,amssymb,wasysym}
\usepackage{dsfont} 
\usepackage{color}
\usepackage[linktocpage=true,colorlinks=true, pdfborder={0 0 0},linkcolor=blue,citecolor=blue,urlcolor=blue,anchorcolor=black,linkcolor=blue]{hyperref}

\newcommand{\op}[1]{\hat{#1}}

\newcommand{\normV}[1]{||{#1}||_2}

\newcommand{\normspec}[1]{||{#1}||_{\text{2}}}

\newcommand{\ident}{\mathds{1}}
\newcommand{\C}{\mathbb{C}}

\newcommand{\quoting}[1]{``#1''}

\newcommand{\rem}[1]{}

\newcommand{\imagc}[1]{\text{Im}\,#1}

\newcommand{\bra}[1]{\langle#1|}
\newcommand{\ket}[1]{|#1\rangle}

\newcommand{\braket}[2]{\langle#1|#2\rangle}

\newcommand{\rca}{\xi}

\newcommand{\Hperturbed}{\op{H}}
\newcommand{\Hp}{\op{H}_1}
\newcommand{\Hcomplement}{\op{H}_{\text{a}}}

\newcommand{\order}{n}

\newcommand{\ev}{E}
\newcommand{\evEP}{\ev_{\text{EP}}}

\newcommand{\state}{R}
\newcommand{\REP}{R_{\text{EP}}}
\newcommand{\LEP}{L_{\text{EP}}}
\newcommand{\stateEP}{\state_{\text{EP}}}

\newcommand{\Jordan}[1]{j_{#1}}

\newcommand{\HEP}{\op{H}_{\text{EP}}}

\newcommand{\ind}{l}
\newcommand{\Jind}{k}
\newcommand{\rigidity}{r}
\newcommand{\PF}{K}
\newcommand{\rcanum}{\rca_{\text{num}}}
\newcommand{\lw}{\gamma}

\newcommand{\HL}[1]{#1}

\begin{document}

\title{Petermann factors and phase rigidities near exceptional points}
\author{Jan Wiersig}
\affiliation{Institut f{\"u}r Physik, Otto-von-Guericke-Universit{\"a}t Magdeburg, Postfach 4120, D-39016 Magdeburg, Germany}
\email{jan.wiersig@ovgu.de}
\date{\today}
\begin{abstract}
The Petermann factor and the phase rigidity are convenient measures for various aspects of open quantum and wave systems, such as the sensitivity of energy eigenvalues to perturbations or the magnitude of quantum excess noise in lasers. We discuss the behavior of these two important quantities near non-Hermitian degeneracies, so-called exceptional points. For small generic perturbations, we derive analytically explicit formulas which reveal a relation to the spectral response strength of the exceptional point. \HL{These formulas shed light on the possibilities for enhanced sensing in passive systems.} The predictions of the general theory are successfully compared to analytical solutions of a toy model. Moreover, it is demonstrated that the connection between the Petermann factor and the spectral response strength provides the basis for an efficient numerical scheme to calculate the latter. 
\HL{Our theory is also important in the presence of the unavoidable imperfections in the fabrication of exceptional points as it allows to determine of what is left of the sensitivity for such imperfect exceptional points studied in experiments.}
\end{abstract}
\maketitle

\section{Introduction}
\label{sec:intro}
Non-Hermitian effective Hamiltonians~\cite{Feshbach58,Feshbach62} describing the dynamics of open quantum and wave systems have attracted considerable interest in recent years, in particular in the context of non-Hermitian photonics~\cite{EMK18,ORN19}. 
One obvious consequence of the non-Hermiticity (or nonself-adjointness) of the Hamiltonian $\Hperturbed \neq \Hperturbed^\dagger$ is that the energy eigenvalues can be complex-valued. The imaginary part has the clear physical interpretation as a decay or growth rate. Another consequence is that if, additionally, the Hamiltonian is nonnormal, $[\Hperturbed,\Hperturbed^\dagger] \neq 0$, then one has to distinguish right eigenstates $\ket{R_\ind}$ from the corresponding left eigenstates~$\ket{L_\ind}$; the quantum number $\ind$ labels the states uniquely. 
The difference of right and left eigenstates is an indicator of the strength of \quoting{non-Hermitian effects}, or better said,  \quoting{nonnormal effects}. This can be made more precise in terms of the Petermann factor of a given pair of eigenstates 
\begin{equation}\label{eq:PF}
\PF_\ind := \frac{\braket{R_\ind}{R_\ind}\braket{L_\ind}{L_\ind}}{|\braket{L_\ind}{R_\ind}|^2}\ ,
\end{equation}
with the conventional inner product $\braket{\cdot}{\cdot}$ in Hilbert space.
There exists a lower bound $\PF_\ind \geq 1$ which follows from the Cauchy-Schwarz inequality $|\braket{L_\ind}{R_\ind}|^2 \leq\braket{R_\ind}{R_\ind}\braket{L_\ind}{L_\ind}$. 
The Petermann factor has been introduced to quantify the linewidth broadening resulting from quantum excess noise in lasers~\cite{Petermann79,Siegman89a,Siegman89b,Siegman95,LEM98,Schomerus09} and laser gyroscopes~\cite{WLY20}. It is therefore also called excess-noise factor or excess-spontaneous emission factor. 
Moreover, the Petermann factor measures the enhanced response of the corresponding eigenstate to perturbations~\cite{HS20}\HL{, which is important for non-Hermitian topological sensors~\cite{BB20} and appears also in the context of the non-Hermitian skin effect, see, e.g., Ref.~\cite{NON23}. In the mathematical literature on nonnormal eigenvalue problems, the enhanced response is equivalently discussed in terms of the condition number (square root of the Petermann factor) in the Bauer-Fike theorem~\cite{BF60} or in terms of the pseudospectrum~\cite{TE05}.}
Even though~$\PF_\ind$ is defined as a quantity characterizing the individual eigenstate with quantum number~$\ind$ it also describes mutual non-orthogonality of eigenstates in two-dimensional Hilbert spaces where the Petermann factors can be written as $\PF_1 = \PF_2 = 1/(1-|\braket{R_1}{R_2}|^2)$ with normalized eigenstates~\cite{LDM97}.

Another but equivalent quantity to characterize the difference of right and left eigenstates is the phase rigidity
\begin{equation}\label{eq:pr}
	\rigidity_\ind :=   \frac{|\braket{L_\ind}{R_\ind}|}{\sqrt{\braket{R_\ind}{R_\ind}\braket{L_\ind}{L_\ind}}} \ ,
\end{equation}
with $0 \leq \rigidity_\ind = 1/{\sqrt{\PF_\ind}} \leq 1$. There are other, slightly different, definitions in the literature. Some agree with Eq.~(\ref{eq:pr}) provided that the normalization of the individual eigenstates are chosen appropriately~\cite{DMX16,XZH19,JBN23}. Some other definitions agree only for (complex-) symmetric Hamiltonians and proper normalization~\cite{LBB97,BRS06,Rotter15,Jin18,ZZJS20}. 
The phase rigidity had been originally invented to quantify the complexness of wavefunctions in systems with partially broken time-reversal symmetry~\cite{LBB97}. \HL{A related quantity has been used to observe signatures of quantum chaos in open billiard systems~\cite{SB05}.}

Of particular interest in the field of open systems are exceptional points (EPs) in parameter space~\cite{Kato66,Heiss00,Berry04,Heiss04,MA19}. 
At such an EP of order $\order$ a variety of interesting phenomena appear. (i) Exactly $\order$ eigenstates and their corresponding eigenvalues of the Hamiltonian coalesce. 
(ii) When the Hamiltonian is subjected to a small perturbation of strength~$\varepsilon > 0$ then the resulting eigenvalue changes are generically proportional to the $\order$th root of~$\varepsilon$~\cite{Kato66}. This feature can be exploited for sensing applications~\cite{Wiersig14b,COZ17,HHW17,XLK19,Wiersig20b,Wiersig20c,KCE22}. The response of the system at the EP in terms of eigenvalue changes can be quantified by the spectral response strength~\cite{Wiersig22}. 
(iii) The overlap $\braket{L_\ind}{R_\ind}$ of each involved eigenstate of the Hamiltonian vanishes~\cite{MF80,Moiseyev11}. The latter implies that the corresponding Petermann factors [Eq.~(\ref{eq:PF})] diverge to infinity~\cite{Berry03,LRS08} and the phase rigidities [Eq.~(\ref{eq:pr})] vanish~\cite{Rotter15} whenever an EP is approached. 
\HL{This has been exploited as an indicator for the appearance of an EP~\cite{DMX16,Jin18,JBN23,XZH19,ZZJS20}.}
The scaling of the Petermann factors near a second-order EP have been studied numerically in Refs.~\cite{Lee09,ZCF10,MFYM20}.
For the phase rigidities such numerical studies have been performed for EPs also of higher order~\cite{DMX16,Jin18,XZH19,ZZJS20}. For a generic perturbation with small perturbation strength~$\varepsilon$, a power law~$\varepsilon^\nu$ with scaling exponent $\nu = (\order-1)/\order$ for the phase rigidities is expected~\cite{Heiss08}.
\HL{The scaling exponent, and in particular deviations from the expected value, have been used to characterize the EP~\cite{DMX16,Jin18,JBN23,XZH19,ZZJS20,Lee09,MFYM20,ZCF10}.}

In this paper, we present a theory for the Petermann factors and the phase rigidities near an EP of arbitrary order. This theory allows to accurately predict their behavior in this extreme situation. In particular, we determine the coefficient in the power law~$\varepsilon^\nu$. We are able to relate this coefficient directly to the spectral response strength of the EP. This relation provides a basis for the calculation of the spectral response strength.

The outline of the paper is as follows. In Sec.~\ref{sec:Preliminaries} some necessary theoretical concepts are reviewed. The scaling properties of the Petermann factors and the phase rigidities near EPs are derived in Sec.~\ref{sec:PFnearEP}. Bounds for these quantities are introduced in Sec.~\ref{sec:bounds}. The results are briefly compared to the literature in Sec.~\ref{sec:literature}. An example and an application are presented in Secs.~\ref{sec:example} and ~\ref{sec:app}. Section~\ref{sec:conclusion} provides a conclusion.

\section{Preliminaries}
\label{sec:Preliminaries}
This section introduces the theoretical concepts necessary for understanding this paper. 

\subsection{Jordan vectors}
\label{sec:jv}
We consider an $\order\times\order$-Hamiltonian $\HEP$ at an EP of order~$\order \geq 2$ with right eigenstate $\ket{\stateEP}$ and eigenvalue $\evEP\in\C$. Clearly, a single state cannot span a basis in the $\order$-dimensional Hilbert space. A basis can be established by the linearly independent (right) Jordan vectors $\ket{\Jordan{1}}, \ket{\Jordan{2}}, \ldots, \ket{\Jordan{\order}}$.
We introduce the $\order\times\order$-matrix
\begin{equation}\label{eq:N}
	\op{N} := \HEP-\evEP\ident 
\end{equation}
which is nilpotent of index $\order$; hence $\op{N}^\order = 0$ but $\op{N}^{\order-1} \neq 0$. The operator~$\ident$ is the identity. With the operator~$\op{N}$ the Jordan chain (see, e.g., Ref.~\cite{SM03}) is defined as
\begin{eqnarray}\label{eq:jc1}
	\op{N}\ket{\Jordan{1}} & = & 0 \ ,\\ 
	\label{eq:jc}
	\op{N}\ket{\Jordan{\Jind}} & = & \ket{\Jordan{\Jind-1}} 
	\;;\, \Jind = 2,\ldots,\order \ .
\end{eqnarray}
It is important to realize that only $\ket{\Jordan{1}} = \ket{\stateEP}$ is a right eigenstate of the Hamiltonian. Note that  Eqs.~(\ref{eq:jc1}) and~(\ref{eq:jc}) do not uniquely determine the Jordan vectors. This can be fixed (up to a complex phase) by requiring~\cite{Wiersig22}
\begin{eqnarray}\label{eq:ortho1}
	\braket{\Jordan{1}}{\Jordan{1}} & = & 1 \ ,\\
	\label{eq:ortho2}
	\braket{\Jordan{\order}}{\Jordan{\Jind}} & = & 0
	\quad\mbox{for}\; \Jind = 1, \ldots, \order-1 \ .
\end{eqnarray}

\subsection{Spectral response strength}
\label{sec:srsEP}
In Ref.~\cite{Wiersig22} the spectral response strength of a system at an EP of order~$\order$ has been determined to be 
\begin{equation}\label{eq:rca}
	\rca = \normspec{\op{N}^{\order-1}}  
\end{equation}
with the $\order\times\order$-matrix $\op{N}$ defined in Eq.~(\ref{eq:N}) and the spectral norm (see, e.g., Ref.~\cite{HJ13})
\begin{equation}\label{eq:defspn}
	\normspec{\op{A}} := \max_{\normV{\psi} = 1}\normV{\op{A}\psi} \ .
\end{equation}
We obey the conventional notation $\normspec{\cdot}$ both for the spectral norm of a matrix [in the left-hand side of Eq.~(\ref{eq:defspn})] and the 2-norm $\normV{\psi} = \sqrt{\braket{\psi}{\psi}}$ of a vector $\ket{\psi}$ [in the right-hand side of Eq.~(\ref{eq:defspn})].
The spectral response strength describes the response of the system at the EP to perturbations,
\begin{equation}\label{eq:H}
	\Hperturbed = \HEP+\varepsilon\Hp \ ,
\end{equation}
in terms of a factor in the bound of the eigenvalue change 
\begin{equation}\label{eq:specresponse}
	|\ev_\ind-\evEP|^\order \leq \varepsilon \normspec{\Hp}\,\rca  \ ,
\end{equation}
where higher orders in the perturbation strength~$\varepsilon$ are ignored.
The scaling of $|\ev_\ind-\evEP|$ with the $\order$th root of~$\varepsilon$ is a manifestation of the enhanced sensitivity of the system at the EP with respect to perturbations.
 \HL{Inequality~(\ref{eq:specresponse}) is valid for generic perturbations, which means here $\op{N}^{\order-1}\Hp\ket{\state_\ind} \neq 0$. In very special situations, parameter variations can correspond to nongeneric perturbations leading to a different scaling behavior. This phenomenon is called  anisotropic EP, see, e.g., Ref.~\cite{XZH19}. It can happen that symmetries lead to nongeneric perturbations, see Refs.~\cite{MB21,SK22}. The presence of a nongeneric perturbation would require to extend the theory presented in Ref.~\cite{Wiersig22} by incorporating the next-order contribution in the Green's function.}

The spectral response strength can be also expressed by the 'length of the last Jordan vector'~\cite{Wiersig22}
\begin{equation}\label{eq:rcajn}
\rca = \frac{1}{\normV{\Jordan{\order}}}
\end{equation}
provided that the normalization and orthogonalization conditions in Eqs.~(\ref{eq:ortho1}) and~(\ref{eq:ortho2}) are applied.

\section{Petermann factors and phase rigidities near an EP}
\label{sec:PFnearEP}
In this section we derive explicit expressions for the Petermann factors [Eq.~(\ref{eq:PF})] and the phase rigidities [Eq.~(\ref{eq:pr})] valid near a given EP of order~$\order$. 

Let us first specify the definition of right and left eigenstates (see, e.g., Ref.~\cite{CM98}) of the Hamiltonian
\begin{equation}\label{eq:biorthogonal}
	\op{H}\ket{R_\ind} = E_\ind\ket{R_\ind}
	\;\;\text{and}\;\;
	\bra{L_\ind}\op{H} = E_\ind\bra{L_\ind} \ .
\end{equation}
There are different possibilities to normalize two of the three inner products $\braket{R_\ind}{R_\ind}$, $\braket{L_\ind}{L_\ind}$, and $\braket{L_\ind}{R_\ind}$. Without loss of generality, we choose the normalization $\braket{R_\ind}{R_\ind} = 1 = \braket{L_\ind}{L_\ind}$ for all $\ind$. In this case the Petermann factors are 
\begin{equation}\label{eq:PFn}
\PF_\ind = \frac{1}{|\braket{L_\ind}{R_\ind}|^2}
\end{equation}
 and the phase rigidities
\begin{equation} \label{eq:prn}
 \rigidity_\ind = |\braket{L_\ind}{R_\ind}| \ . 
\end{equation} 
 The central quantity to compute is therefore $|\braket{L_\ind}{R_\ind}|$.

\begin{figure}[ht]
	\includegraphics[width=0.95\columnwidth]{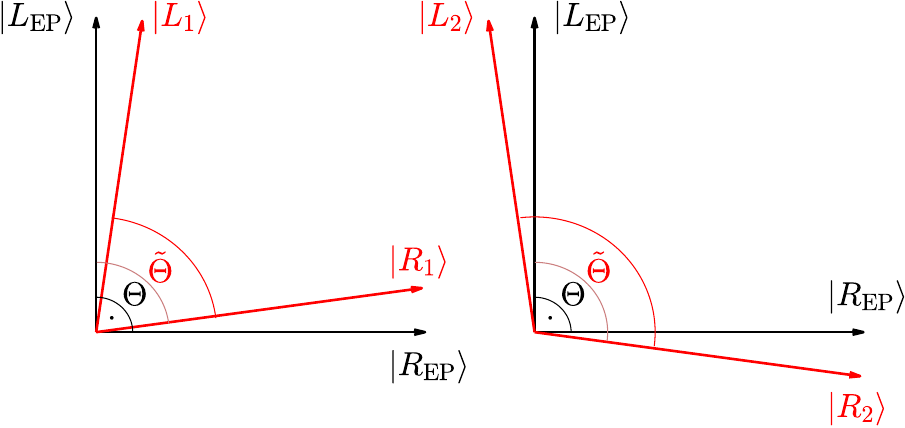}
	\caption{\HL{Illustration of Eq.~(\ref{eq:ntimes}) for $\order = 2$ and normalized real-valued vectors. With $\braket{L_\ind}{R_\ind} = \cos{\tilde{\Theta}}$, $\braket{\LEP}{R_\ind} = \cos{\Theta}$, and $\braket{\stateEP}{\LEP} = 0$ holds $\tilde{\Theta} < \Theta < \pi/2$ (left) and $\tilde{\Theta} > \Theta > \pi/2$ (right). Both cases are consistent with Eq.~(\ref{eq:ntimes}) corresponding to $\cos{\tilde{\Theta}} = 2\cos{\Theta}$.}}
	\label{fig:LR}
\end{figure}

We start with the perturbed Hamiltonian~(\ref{eq:H}) and add another perturbation $\Delta\varepsilon\Hp$ with $0 < \Delta\varepsilon \ll \varepsilon$. For what follows it is crucial that the two perturbations can be combined to $(\varepsilon+\Delta\varepsilon)\Hp$. The change in the eigenvalues can be evaluated in first-order non-Hermitian perturbation theory~\cite{BB20}
\begin{equation}\label{eq:firstorderPT}
\Delta E_\ind = \Delta\varepsilon\frac{\bra{L_\ind}\Hp\ket{R_\ind}}{\braket{L_\ind}{R_\ind}} \ .	
\end{equation}
In the limit $\Delta\varepsilon \to 0$ we can write
\begin{equation}\label{eq:LR}
\braket{L_\ind}{R_\ind} = \left(\frac{dE_\ind(\varepsilon)}{d\varepsilon}\right)^{-1}\bra{L_\ind}\Hp\ket{R_\ind} \ .
\end{equation}
For a generic perturbation the eigenvalue change is $E_\ind-\evEP = \varepsilon^{1/\order}e_\ind$ with $e_\ind\in\C$~\cite{Kato66}. The $\ind$-dependence of $e_\ind$ originates from the primitive $\ind$th root of unity $\exp{(i2\pi \ind/\order)}$ with $\ind = 1,\ldots,\order$.
We deduce that 
\begin{equation}\label{eq:dedeps}
\frac{dE_\ind}{d\varepsilon} = \frac{\varepsilon^{\frac{1}{\order}-1}e_\ind}{\order} \ . 
\end{equation}	
Moreover, in leading order we can approximate $\bra{L_\ind}\Hp\ket{R_\ind}$ by
\begin{equation}\label{eq:LEPR}
\bra{\LEP}\Hp\ket{R_\ind} = \frac{1}{\varepsilon} (E_\ind-\evEP)\braket{\LEP}{R_\ind}
\end{equation}
and therefore with Eqs.~(\ref{eq:LR}) and~(\ref{eq:dedeps})
\begin{equation}\label{eq:ntimes}
\braket{L_\ind}{R_\ind} = \order\braket{\LEP}{R_\ind} \ .
\end{equation}
\HL{This equation is illustrated in Fig.~\ref{fig:LR}.}

To calculate the right-hand side of Eq.~(\ref{eq:ntimes}) we first expand $\ket{R_\ind}$ in terms of the Jordan vectors
\begin{equation}\label{eq:Rexpansion}
\ket{R_\ind} = \sum_{\Jind=1}^\order\alpha^{(\ind)}_\Jind\ket{\Jordan{\Jind}} \ .
\end{equation}
In conjunction with $\op{N}^{\order-1}\ket{\Jordan{\order}} = \ket{\Jordan{1}} = \ket{\REP}$ and $\op{N}^{\order-1}\ket{\Jordan{\Jind}} = 0$ for $\Jind < \order$ [see the Jordan chain in Eqs.~(\ref{eq:jc1}) and (\ref{eq:jc})] we get 
\begin{equation}\label{eq:alpha0}
\alpha^{(\ind)}_\order = \bra{\REP}\op{N}^{\order-1}\ket{R_\ind} \ .
\end{equation}
In Ref.~\cite{Wiersig22c} it has been shown that generically for small perturbations
\begin{equation}
(E_\ind-\evEP)^\order = \bra{R_\ind}\op{N}^{\order-1}\varepsilon\Hp\ket{R_\ind} \ .
\end{equation}
In the leading order we replace $\ket{R_\ind}$ by $\ket{\REP}$ and back resulting with the help of Eq.~(\ref{eq:H}) in 
\begin{eqnarray}
\nonumber
(E_\ind-\evEP)^\order & = & \bra{R_\ind}\op{N}^{\order-1}\Hperturbed\ket{R_\ind}  -\bra{R_\ind}\op{N}^{\order-1}\HEP\ket{\REP} \\
& = & (E_\ind-\evEP)\bra{\REP}\op{N}^{\order-1}\ket{R_\ind} \ .
\end{eqnarray}
In combination with Eq.~(\ref{eq:alpha0}) we then get
\begin{equation}\label{eq:alpha}
	\alpha^{(\ind)}_\order = (E_\ind-\evEP)^{\order-1} \ .
\end{equation}
Next, we take advantage of the relation proven in Appendix~\ref{app:LEPJordan}
\begin{equation}\label{eq:LEPJordan}
|\braket{\LEP}{\Jordan{\Jind}}| = \normV{\Jordan{\order}}\delta_{\Jind\order} 
\end{equation}
with Kronecker delta $\delta_{\Jind\order}$. Together with Eqs.~(\ref{eq:ntimes}) and (\ref{eq:Rexpansion}) we obtain 
\begin{equation}
|\braket{L_\ind}{R_\ind}| = \order 	|\alpha^{(\ind)}_\order|\normV{\Jordan{\order}} \ .
\end{equation}	
With Eqs.~(\ref{eq:rcajn}), (\ref{eq:prn}), and~(\ref{eq:alpha}) we get our central result for the phase rigidities
\begin{equation}\label{eq:rigidityresult}
\rigidity_\ind = \frac{\order|E_\ind-\evEP|^{\order-1}}{\rca} \ .
\end{equation}
Since the absolute value of the eigenvalue change in the leading order is independent of the quantum number~$\ind$~\cite{Wiersig22c} we conclude that the same is true for the phase rigidities. Hence, in the following we speak about the phase rigidity $\rigidity = \rigidity_\ind$ in the vicinity of the EP. Of course, this also applies to the Petermann factor(s) [Eq.~(\ref{eq:PFn})]
\begin{equation}\label{eq:PFresult}
	\PF_\ind = \frac{\rca^2}{\order^2|E_\ind-\evEP|^{2\order-2}} = \PF \ .
\end{equation}
In contrast to inequality~(\ref{eq:specresponse}), Eqs.~(\ref{eq:rigidityresult}) and~(\ref{eq:PFresult}) are really equations. Hence, the spectral response strength~$\rca$ here does not just describe a bound but an exact relationship. 

At first glance, it might be surprising that the Petermann factor and the phase rigidity are related to the spectral response strength and not to the eigenstate response strength defined in Ref.~\cite{Wiersig22}. After all, the Petermann factor and the phase rigidity have been introduced as measures of the eigenstates. But both quantities are also related to spectral properties via the first-order non-Hermitian perturbation theory as visible in Eq.~(\ref{eq:firstorderPT}).

\section{Bounds}
\label{sec:bounds}
We combine Eqs.~(\ref{eq:rigidityresult}) and~(\ref{eq:PFresult}) with inequality~(\ref{eq:specresponse}) leading to the bounds
\begin{equation}\label{eq:rigiditybound}
	\rigidity \leq \frac{\order \left(\varepsilon \normspec{\Hp}\right)^{\frac{\order-1}{\order}}}{\rca^{\frac{1}{\order}}} 
\end{equation}
and
\begin{equation}\label{eq:PFbound}
	\PF \geq \frac{\rca^{\frac{2}{\order}}}{\order^2 \left(\varepsilon \normspec{\Hp}\right)^{\frac{2\order-2}{\order}}} \ .
\end{equation}
These inequalities provide easy-to-calculate bounds for the phase rigidity and the Petermann factor provided that the spectral response strength of the EP is known and the size of the perturbation $\normspec{\Hp}$ can be calculated or at least estimated.  

For passive (no gain) systems Ref.~\cite{Wiersig22b} has derived an upper bound for the spectral response strength 
\begin{equation}\label{eq:rcapassve}
	\rca \leq \left(\sqrt{2\order}|\imagc{\evEP}|\right)^{\order-1} \ .
\end{equation}
Based on this inequality we derive an upper bound for the Petermann factor in Eq.~(\ref{eq:PFresult})
\begin{equation}\label{eq:PFpassive0}
\PF \leq 2^{\order-1}\order^{\order-3}\left(\frac{|\imagc{\evEP}|}{|E_\ind-\evEP|}\right)^{2\order-2} \ .
\end{equation}
An analog inequality for the phase rigidity can be easily derived. 

The inequality~(\ref{eq:PFpassive0}) has an important consequence. Assume that we want to utilize our system as a sensor but in contrast to a conventional EP-based sensor where the energy (or frequency) splitting is detected we want to spoil the EP before the detection such that we have isolated peaks with linewidth of roughly $\lw = 2|\imagc{\evEP}|$ in the spectrum. We want to carry out the sensing with one of these isolated peaks taking advantage of the enhanced sensitivity of the corresponding eigenvalue expressed by $\PF > 1$. To resolve the individual peaks experimentally by standard means we need a splitting $2|E_\ind-\evEP|$ (approximate maximum distance between two eigenvalues along the real axis, valid for a not too small order~$\order$) at least of the size of the linewidth $\lw$. Inserting this into inequality~(\ref{eq:PFpassive0}) yields
\begin{equation}\label{eq:PFpassive}
	\PF \leq  2^{\order-1} \order^{\order-3} \ .
\end{equation}
For $\order = 2$ the right-hand side equals unity, hence the isolated peaks cannot be used for enhanced sensing. However, for $\order = 2$ the estimation of the splitting being $2|E_\ind-\evEP|$ is not a good one anyway since, in the worst case, the induced energy change may be only in the imaginary part. For $\order > 2$ this is impossible, so the estimation of the splitting makes sense in that case. Here, the right-hand side of the inequality~(\ref{eq:PFpassive}) is~$\geq 4$ and increases strongly with increasing $\order$. Therefore, for orders~$\order \geq 3$ the isolated peaks can be exploited for enhanced sensing. 
Note that there is no limitation at all for systems with gain. 

\section{Comparison to literature}
\label{sec:literature}
Having in mind that for a generic perturbation with sufficiently small strength~$\varepsilon$ the eigenvalue change scales as $E_\ind-\evEP \propto \varepsilon^{1/\order}$, it is clear that our result in Eq.~(\ref{eq:rigidityresult}) is in full agreement with the expected power law $\propto\varepsilon^\nu$ with $\nu = (\order-1)/\order$ for generic perturbations~\cite{Heiss08}. Noteworthy, our approach goes well beyond that of Ref.~\cite{Heiss08} as we have explicitly determined the coefficient in front of the power law. In addition, we have related this coefficient to the spectral response strength of the EP.

The numerical studies in the literature~\cite{DMX16,Jin18,JBN23,XZH19,ZZJS20} have utilized the vanishing of the phase rigidity as an indicator for the appearance of an EP. Once an EP is found, the scaling behavior of the phase rigidity has been employed to characterize the EP. Often the power law with generic scaling exponent $\nu = (\order-1)/\order$ has been confirmed but sometimes other exponents have been obtained. Such deviating exponents indicate parameter variations that belong to nongeneric perturbations of the Hamiltonian. 
For example, Ref.~\cite{DMX16} has reported $\nu = 1$ for a second-order EP and  $\nu = 3/4$ for a fourth-order EP. The former deviates from the generic behavior and the latter agrees with it. The same for Ref.~\cite{JBN23} where $\nu = 1$ for $\order = 2$ and $\nu = 2/3$ for $\order = 3$ has been observed. Ref.~\cite{Jin18} found $\nu = 3/4$ for $\order = 4$ consistent with the generic behavior.
For so-called anisotropic EPs with nongeneric eigenvalue changes usually nongeneric scaling of the phase rigidity has been observed, e.g. with exponents $\nu = (\order-1)/2$ and $\order-1$~\cite{XZH19}. Similarly, Ref.~\cite{ZZJS20} has reported the scaling exponent $\nu = (\order-1)/2$ for supersymmetric arrays. Also here the perturbation is not generic leading to a square-root eigenvalue change for the EP even for $\order 
> 2$.

With $\PF = 1/\rigidity^2$ the expected scaling behavior for the Petermann factor is $\varepsilon^{-2\nu}$ with $\nu = (\order-1)/\order$ which is in full agreement with our result in Eq.~(\ref{eq:PFresult}). The numerical studies of the Petermann factor for a one-dimensional barrier~\cite{Lee09}, a disordered dimer chain~\cite{ZCF10}, and a ring of coupled-dimer cavities with embedded parity-time symmetric defects~\cite{MFYM20} show an $1/\varepsilon$-scaling near a second-order EP which is consistent with our general result.

\section{Example}
\label{sec:example}
We consider a simple model system where explicit results can be obtained enabling us to see clearly the validity of our general theory introduced in the previous section. The Hamiltonian of the unperturbed system is 
\begin{equation}\label{eq:HNmodel}
	\HEP = \left(\begin{array}{ccccc}
		E_0    & A    & 0       & \ldots & 0  \\
		0      & E_0    & A     & \ldots & 0  \\
		0      & 0      & E_0     & \ldots & 0  \\
		\vdots & \vdots & \vdots  &  \ddots       & \vdots   \\
		0      & 0 & 0       &  \ldots     & E_0\\
	\end{array}\right) \ .
\end{equation}
This $\order\times\order$ Hamiltonian is a nonperiodic, fully asymmetric limiting case of the Hatano-Nelson model~\cite{HN96}. It describes a directed hopping between nearest neighbors in a tight-binding chain. For a nonzero hopping parameter~$A\in\C$, the Hamiltonian is at an EP of order $\order$ with eigenvalue $\evEP = E_0$. The spectral response strength has been calculated in Ref.~\cite{Wiersig22} to be
\begin{equation}\label{eq:HNM1}
	\rca = |A|^{\order-1} \ .
\end{equation}
We consider the perturbation in Eq.~(\ref{eq:H}) with  
\begin{equation}\label{eq:HNpert}
	\Hp = \left(\begin{array}{ccccc}
		0      & 0      & 0       & \ldots & 0  \\
		0      & 0      & 0     & \ldots & 0  \\
		0      & 0      & 0     & \ldots & 0  \\
		\vdots & \vdots & \vdots  &  \ddots       & \vdots   \\
		1      & 0 & 0       &  \ldots     & 0\\
	\end{array}\right) \ .
\end{equation}
This perturbation is simple enough to be dealt with analytically but still it is generic in the sense that it leads to an $\varepsilon^{1/\order}$-scaling of the energy eigenvalue change. In fact, a short calculation shows
\begin{equation}\label{eq:HNsplitting}
\ev_\ind-\evEP = \varepsilon^{\frac{1}{\order}}A^{\frac{\order-1}{\order}} \ .
\end{equation}
Together with Eq.~(\ref{eq:HNM1}) we obtain
\begin{equation}\label{eq:HNsplitting2}
	|\ev_\ind-\evEP| = (\varepsilon\rca)^{\frac{1}{\order}} \ .
\end{equation}
This result is consistent with inequality~(\ref{eq:specresponse}) since the spectral norm of the perturbation in Eq.~(\ref{eq:HNpert}) is $\normspec{\Hp} = 1$.

\begin{figure}[ht]
	\includegraphics[width=0.85\columnwidth]{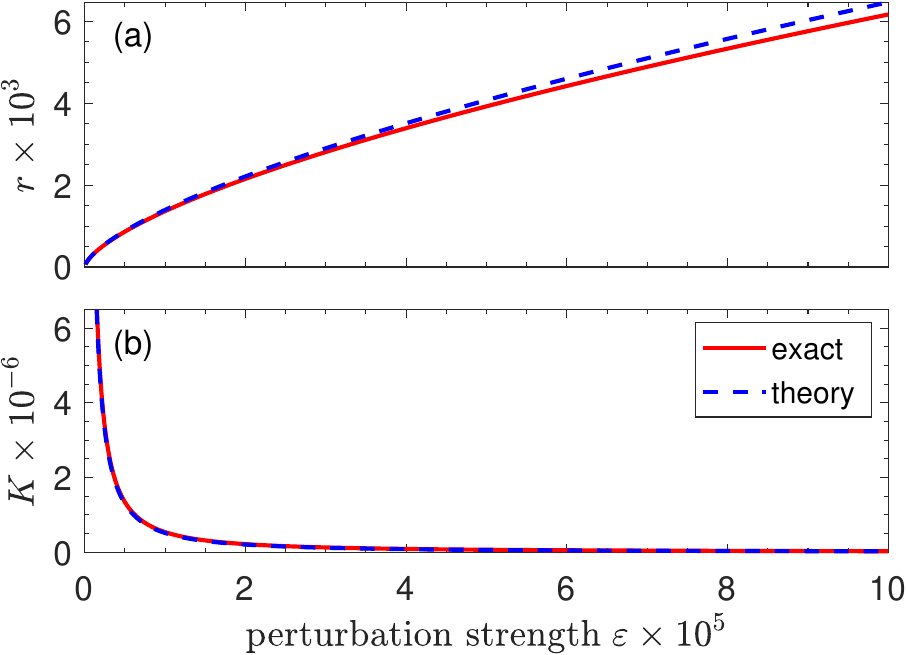}
	\caption{(a) Phase rigidity $\rigidity$ and (b) Petermann factor $\PF$ vs perturbation strength $\varepsilon$ (all quantities are dimensionless) for the perturbed hopping model in Eqs.~(\ref{eq:HNmodel}) and~(\ref{eq:HNpert}). The solid red curve shows the exact result in Eq.~(\ref{eq:HNr}) [with Eq.~(\ref{eq:HNsplitting})] and the dashed blue curve the theoretical result in Eq.~(\ref{eq:HNrapprox}) valid for small~$\varepsilon$. In (b) the curves lie on top of each other. The parameters are $A = 1$, $E_0 = 0$, and $\order = 3$.}
	\label{fig:Hatano}
\end{figure}

A more tedious but also straightforward calculation of the inner product $\braket{L_\ind}{R_\ind}$ yields for the phase rigidity
\begin{equation}\label{eq:HNr}
\rigidity = \frac{\order\left|\frac{\ev_\ind-\evEP}{A}\right|^{\order-1}}{\sum_{j = 1}^{\order}\left|\frac{\ev_\ind-\evEP}{A}\right|^{j-1}} \ .
\end{equation}
For small perturbation strength $\varepsilon$ the eigenvalue change in Eq.~(\ref{eq:HNsplitting}) is small, therefore in Eq.~(\ref{eq:HNr}) only the $j = 1$-term contributes, and consequently the denominator goes to unity. Hence, with Eq.~(\ref{eq:HNM1}) follows
\begin{equation}\label{eq:HNrapprox}
\rigidity =\frac{\order|E_\ind-\evEP|^{\order-1}}{\rca} \ ,
\end{equation}
in accordance with the general theoretical result in Eq.~(\ref{eq:rigidityresult}). Figure~\ref{fig:Hatano} shows a comparison of the exact result with the theoretical result for the phase rigidity $\rigidity$ and the Petermann factor $\PF = 1/\rigidity^2$. For small perturbation the agreement is nearly perfect. 

Note that for this simple example the bounds in the inequalities~(\ref{eq:rigiditybound}) and~(\ref{eq:PFbound}) agree with the theoretical result in Eqs.~(\ref{eq:rigidityresult}) and~(\ref{eq:PFresult}).

\section{Application}
\label{sec:app}
The computation of the spectral response strength $\rca$ of an EP of arbitrary order~$\order$ in Eq.~(\ref{eq:rca}) is easy and can be performed most of the time by hand. However, the equation requires the Hamiltonian to be an $\order\times\order$-matrix, which is a severe limitation. The same is true for the alternative computation based on the last Jordan vector in Eq.~(\ref{eq:rcajn}). For the more general case of an EP of order~$\order$ embedded in an Hilbert space of dimension $m > \order$ no method of computation has been presented yet. 

\HL{It is natural to assume that the revealed connection of $\PF_\ind$ and $\rigidity_\ind$ with the spectral response strength~$\rca$ is a local property, even though we cannot provide an analytical proof here. This would offer a strategy for the computation of~$\rca$ in the higher dimensional case. 
Let us consider an $m\times m$-Hamiltonian with an EP with order $\order < m$ and eigenvalue~$\evEP$. We perturb this system slightly by a randomly selected perturbation, which is with probability 1 a generic perturbation. The strength of which is a parameter of the method and has to be chosen to be large enough to move the system away from the EP. Otherwise $\PF_\ind$ would diverge and $\rigidity_\ind$ vanishes and therefore Eqs.~(\ref{eq:rigidityresult}) and~(\ref{eq:PFresult}) could not be used. At the same time the perturbation strength should be small enough such that the leading-order approximation which led to Eqs.~(\ref{eq:rigidityresult}) and~(\ref{eq:PFresult}) is still valid. Note that  already the numerical implementation, even if double-precision floating-point arithmetic is used, can drive the system significant away from a higher-order EP~\cite{Wiersig22c}. In such a case no additional random perturbation is needed.}
For simplicity we assume that no other EP with eigenvalue close to $\evEP$ exists. The method starts with computing all eigenvalues~$\ev_\ind$, right eigenstates $\ket{R_\ind}$, and left eigenstates~$\ket{L_\ind}$ of the Hamiltonian by a standard numerical routine. 
\HL{Next, the eigenstates and eigenvalues that are associated with the desired EP are selected. To do so, we determine from the set of eigenstates with $|\braket{L_\ind}{R_\ind}|$ small the quantum number~$\ind$ for which the deviation $|\ev_\ind-\evEP|$ is minimal.} This minimum energy deviation we denote as~$\Delta\ev$.  Finally, we plug $\Delta\ev$ and the corresponding phase rigidity $\rigidity = |\braket{L_\ind}{R_\ind}|$ into the equation for the numerically determined spectral response strength
\begin{equation}
\rcanum = \frac{\order\Delta\ev^{\order-1}}{\rigidity} \ ,
\end{equation}
which has been obtained from Eq.~(\ref{eq:rigidityresult}).

We demonstrate in the following that this simple scheme is quite efficient. To produce many numerical examples, we adopt the random-matrix approach invented in Ref.~\cite{Wiersig22}. We introduce the $\order\times\order$ matrix $\HEP$ having an EP with order~$\order$ and eigenvalue $\evEP$ via a similarity transformation $\HEP = \op{Q}\op{J}\op{Q}^{-1}$, with $\op{J}$ being an $\order\times\order$ matrix with an EP of order~$\order$ in Jordan normal form and $\op{Q}$ is an, in general nonunitary, $\order\times\order$ matrix consisting of complex random numbers with real and imaginary parts being drawn from a uniform distribution on the interval $[-\frac{1}{2},\frac{1}{2}]$. We calculate the associated spectral response strength~$\rca$ simply from Eq.~(\ref{eq:rca}) and save the value for later comparison. 
Subsequently, we create the $(m-\order)\times(m-\order)$-matrix $\Hcomplement$ with its elements to be complex random numbers with real and imaginary parts being drawn from a uniform distribution on $[-\frac{1}{2},\frac{1}{2}]$. 
In the end, we combine these two matrices to obtain the $m\times m$ Hamiltonian
\begin{equation}\label{eq:combinedH}
\op{H} = \op{U}\left(\begin{array}{cc}
\HEP & 0\\
0 & \Hcomplement \\
\end{array}\right)\op{U}^\dagger \ .
\end{equation}
The random unitary matrix~$\op{U}$ is generated with the help of a QR decomposition of a random complex $m\times m$-matrix~\cite{Mezzadri07} constructed in the same way as $\Hcomplement$ above. \HL{The resulting Hamiltonian~$\op{H}$ has an EP with the same eigenvalue $\evEP$, the same spectral response strength~$\rca$, and the same order~$\order$ than the original Hamiltonian $\HEP$. We find that the small numerical errors in the above procedure (double-precision floating-point arithmetic in MATLAB is used) drive $\op{H}$ slightly away from the EP. Hence, no additional random perturbation is needed here.} The Hamiltonian~$\op{H}$ is the input of the proposed scheme to compute the spectral response strength $\rcanum$ of the EP. 

\begin{figure}[ht]
	\includegraphics[width=0.95\columnwidth]{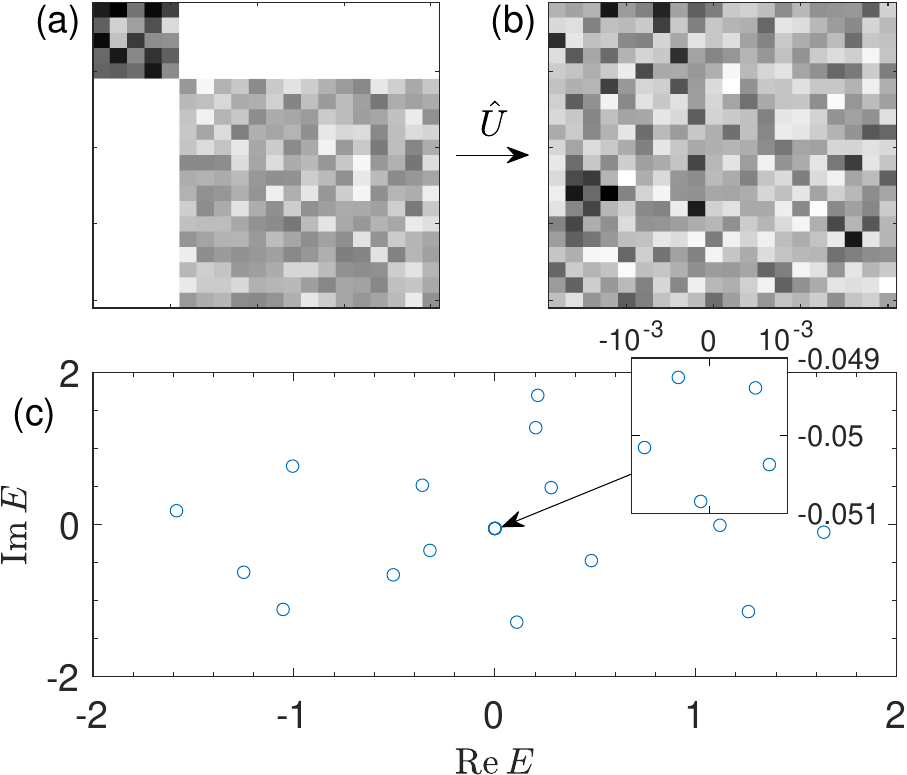}
	\caption{Illustration of the construction of a random $20\times 20$-Hamiltonian possessing a (slightly perturbed) EP of order 5 with eigenvalue $\evEP = -i0.05$. (a) and (b) show a realization of the Hamiltonian in Eq.~(\ref{eq:combinedH}) without and with the random unitary matrix~$\op{U}$, respectively. The absolute values of the matrix elements are plotted in a grayscale representation: The maximum value is black and the minimum value of zero is shown in white. (c) depict the complex eigenvalues (dimensionless) of the Hamiltonian in Eq.~(\ref{eq:combinedH}). Inset: Magnification around the slightly perturbed EP.}
	\label{fig:randomEP}
\end{figure}

Using MATLAB the proposed procedure is illustrated in Fig.~\ref{fig:randomEP}. In Fig.~\ref{fig:randomEP}(a) the Hamiltonian in Eq.~(\ref{eq:combinedH}) is shown before the random unitary matrix~$\op{U}$ is applied. In the colormap the submatrix $\HEP$ appears darker than the submatrix~$\Hcomplement$ because the former originates from the matrix $\op{J}$ where the matrix elements are 0 or 1 whereas the latter has random matrix elements in $[-\frac{1}{2},\frac{1}{2}]$. Figure~\ref{fig:randomEP}(b) depicts the Hamiltonian~$\op{H}$ in Eq.~(\ref{eq:combinedH}) after the unitary transformation. The Hamiltonian looks random but it still possesses the EP inherited from~$\HEP$ with the same spectral response strength (up to numerical uncertainties).

Figure~\ref{fig:randomEP}(c) shows the eigenvalues of~$\op{H}$. The magnification in the inset reveals that the EP is slightly perturbed due to numerical uncertainties which lifts the degeneracy and results into a ring of five eigenvalues. This small splitting is essential for the proposed numerical scheme. 
As a side remark it is briefly mentioned that the appearance of the splitting of EP eigenvalues due to numerical uncertainties or fabrication imperfection is reminiscent to the vortex-splitting phenomenon~\cite{RLE12} due to the instability of higher-order optical vortices.

\begin{figure}[ht]
	\includegraphics[width=0.95\columnwidth]{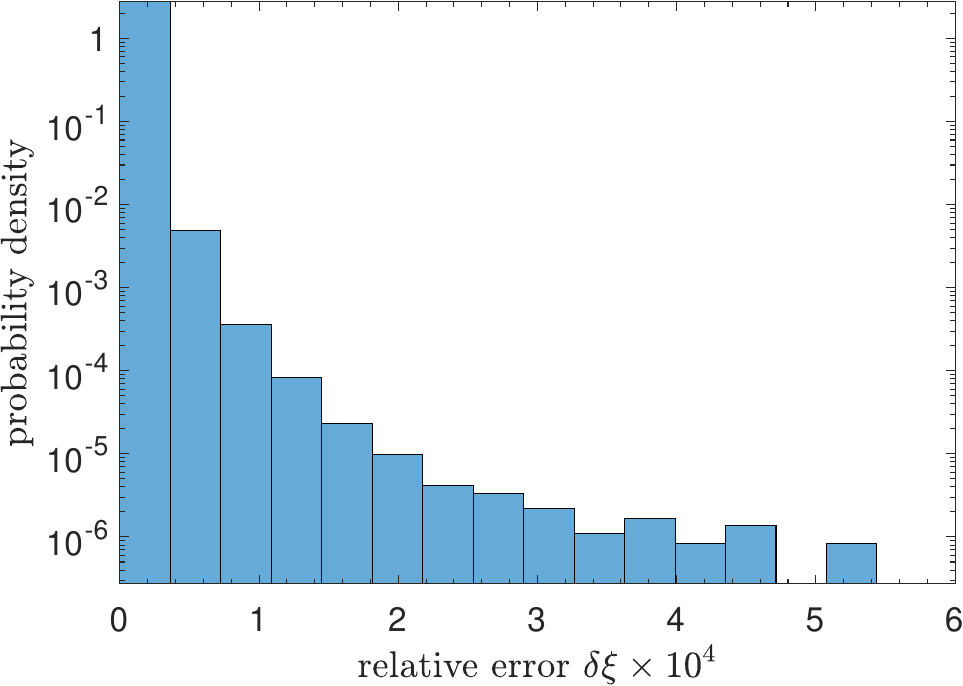}
 	\caption{Probability density function of the dimensionless relative error $\delta\rca := |\rcanum-\rca|/\rca$ computed from $10^7$ realizations of random $20\times 20$ Hamiltonians very close to a third-order EP with eigenvalue~$\evEP = -i0.05$. A few outliers outside the depicted interval, but with $\delta\rca < 11\times 10^{-4}$, are not shown. Note the logarithmic scale on the vertical axis.}
 	\label{fig:histoPF}
 \end{figure}

Figure~\ref{fig:histoPF} shows such a comparison for as many as $10^7$ realizations of $20\times 20$ Hamiltonians~$\op{H}$ each very close to an EP of third order. The relative error $|\rcanum-\rca|/\rca$ is below one-tenth of a percent. The raw data (not shown) unveil that the largest deviations occur whenever the eigenvalue splitting due to the finite machine precision is relatively pronounced. Here the leading-order contribution may not always be sufficient for calculating the phase rigidity accurately enough. This happens in particular for larger order~$\order$ \HL{where small perturbations lead to large splittings. We observe relative errors around and below $1\%$ for $\order < 5$ (not shown). 
Our numerics demonstrate that the connection between phase rigidity and spectral response strength can be exploited to compute the latter even for the case where the dimension of the Hilbert space is larger than the order of the EP.}
 
The introduced scheme for the computation of the spectral response strength~$\rca$ could possibly be also applied to systems where no (effective) Hamiltonian is available but instead a wave equation, e.g., for optical microcavities~\cite{Wiersig02b}\HL{, where the Petermann factors can be directly calculated from the spatial mode structure~\cite{Schomerus09}}. Based on the knowledge gained from the presented theory, we can interpret the fitting procedure in Ref.~\cite{Wiersig16} to determine numerically the internal backscattering coefficient in a perturbed microdisk system (the absolute value is $\rca$ in this context) near a second-order EP as a special case of our scheme\HL{; the same for the recently published fitting procedure in Ref.~\cite{KGY23}}. Once again, it is emphasized that our approach is much more general as it applies to a wide range of systems and EPs of arbitrary order. 
A good test system for applying the above scheme would be the situation in Refs.~\cite{KYW18,KW19} where both a wave equation and an effective Hamiltonian are available simultaneously.

\section{Conclusion}
\label{sec:conclusion}
We have derived explicit formulas for the Petermann factors and phase rigidities near EPs of arbitrary order. Our results not only confirm the known scaling behavior for generic perturbations but also provide the prefactor. Our theory has revealed an unexpected relation to the spectral response strength of the EP. 
Bounds for the Petermann factors and phase rigidities have been introduced that can be used for an easy estimation of the order of magnitude and in the case of passive systems allow for an assessment of the possibilities for enhanced sensing.
We have discussed our results in the context of numerical studies in the literature. %
The predictions of our general theory have been compared to the solutions of an analytically solvable model and very good agreement has been observed.

We have demonstrated that the established connection of Petermann factors and phase rigidities with the spectral response strength can be exploited for the numerical computation of the latter even in the case of a higher-dimensional Hilbert space. The method is efficient and reliable for EPs with order up to five \HL{for double-precision floating-point arithmetic. EPs with even higher order are so sensitive to numerical imperfections that the highest-order contribution of the Green's function is not always sufficient.} 

Realizing EPs experimentally is not an easy task in particular for EPs of higher order because of their extreme sensitivity to fabrication imperfections. In practice, the experimental setup is never located exactly on the EP but at best in its vicinity. Our theory of the Petermann factors and phase rigidities is able to predict of what is left of the sensitivity for such imperfect EPs. This is a valuable information for sensing applications.
This applies in particular to experiments which are deliberately detuned from the EP to  mitigate the harmful effects of quantum excess noise~\cite{KCE22}.

Moreover, our results can be beneficial for the study of extreme dynamics near higher-order EPs~\cite{ZCK18}.

\acknowledgments 
Fruitful discussions with M.~P. van Exter \HL{and H. Schomerus} are acknowledged. 

\begin{appendix}
\section{Computation of $|\braket{\LEP}{\Jordan{\Jind}}|$}
\label{app:LEPJordan}
This Appendix contains a short proof of Eq.~(\ref{eq:LEPJordan}). For the case $\Jind < \order$ we can write according to Eq.~(\ref{eq:jc})
\begin{equation}
	\braket{\LEP}{\Jordan{\Jind}} = \bra{\LEP}\op{N}\ket{\Jordan{\Jind+1}} = \left(\bra{\LEP}\op{N}\right)\ket{\Jordan{\Jind+1}}\ .
\end{equation}
With the definition of the operator $\op{N}$ in Eq.~(\ref{eq:N}) it is clear that
\begin{equation}
\bra{\LEP}\op{N} = \bra{\LEP}\HEP - \bra{\LEP}\evEP = 0 \ .
\end{equation}
Hence, 
\begin{equation}\label{eq:applj}
\braket{\LEP}{\Jordan{\Jind}} = 0 \quad\text{for}\;\Jind < \order \ .
\end{equation}
In words, $\ket{\LEP}$ is orthogonal to all Jordan vectors $\ket{\Jordan{\Jind}}$ for $\Jind < \order$. The last Jordan vector $\ket{\Jordan{\order}}$ is also  orthogonal to all other Jordan vectors $\ket{\Jordan{\Jind}}$ because of the chosen orthogonalization conditions in Eq.~(\ref{eq:ortho2}). We conclude that the unit vector $\ket{\LEP}$ is parallel (up to a complex phase) to $\ket{\Jordan{\order}}$ and hence $|\braket{\LEP}{\Jordan{\order}}| = \normV{\Jordan{\order}}$. Together with Eq.~(\ref{eq:applj}) this proves Eq.~(\ref{eq:LEPJordan}). 
\end{appendix}

%

\end{document}